\documentclass[10pt,english]{article}
\usepackage[T1]{fontenc}
\usepackage[latin9]{inputenc}
\usepackage[a4paper]{geometry}
\usepackage{float}
\usepackage{amsmath}
\usepackage{amssymb}
\usepackage{graphicx}
\usepackage{setspace}
\usepackage{wasysym}
\usepackage{esint}
\makeatletter
\newcommand{\lyxaddress}[1]{
	\par {\raggedright #1
	\vspace{1.4em}
	\noindent\par}
}

\makeatother

\usepackage{babel}
\begin{document}
\title{Pseudoscalar Meson Mixing, the Contribution of the Hadronic Continuum
to Deviation from Factorization}
\author{N. F. Nasrallah}
\maketitle

\lyxaddress{\begin{center}
Faculty of Science, Lebanese University. Tripoli 1300, Lebanon
\par\end{center}}
\begin{abstract}
The contribution of the hadronic continuum in the QCD sum rule calculation
of the parameters entering in pseudoscalar meson mixing is evaluated
by making use of simple integration kernels tailored in order to practically
eliminate the contribution of the hadronic continuum. This approach
avoids the arbitrariness and instability inherent to previous sum
rule calculations. An independent evaluation of the mixed quark gluon
condensate $\left\langle QGC\right\rangle =$$\left\langle g\bar{q}\sigma_{\mu\nu}\frac{\lambda^{a}}{2}G_{\mu\nu}^{a}q\right\rangle $
which enters in the calculation is presented as well as the calculation
of the K-meson decay constant $f_{K}$ to five loops. 
\end{abstract}

\section{Introduction}

A considerable amount of attention \cite{Nierste} has been devoted
to the study of neutral meson mixing. In particular the matrix elements
$\left\langle \bar{K^{0}}(p')\left|\Theta_{\Delta S=2}\right|K^{0}(p)\right\rangle $\cite{Multi1}
and $\left\langle \bar{B^{0}}(p')\left|\Theta_{\Delta B=2}\right|B^{0}(p)\right\rangle $
\cite{Koerner,Aoki}

where $\Theta_{\Delta S=2}=(\bar{s}_{L}\gamma_{\mu}d_{L})(\bar{s}_{L}\gamma_{\mu}d_{L})$
and $\Theta_{\Delta B=2}=(\bar{b}_{L}\gamma_{\mu}d_{L})(\bar{b}_{L}\gamma_{\mu}d_{L})$
which contribute to the mass differences of the neutral mesons and
in studies of CP violation have been classic subjects of investigation.
The simplest approach (factorization) reduces these matrix elements
to the products

\begin{align}
\left\langle \bar{K^{0}}\left|\Theta_{\Delta S=2}\right|K^{0}\right\rangle  & =\left\langle \bar{K^{0}}\left|\bar{s}_{L}\gamma_{\mu}d_{L}\right|0\right\rangle \left\langle 0\left|\bar{s}_{L}\gamma_{\mu}d_{L}\right|K^{0}\right\rangle 
\end{align}

\begin{equation}
\left\langle \bar{B^{0}}\left|\Theta_{\Delta B=2}\right|B^{0}\right\rangle =\left\langle \bar{B^{0}}\left|\bar{b}_{L}\gamma_{\mu}d_{L}\right|0\right\rangle \left\langle 0\left|\bar{b}_{L}\gamma_{\mu}d_{L}\right|B^{0}\right\rangle 
\end{equation}

deviation from factorization is described by a parameter $B$ which
multiplies the above matrix elements. In factorization $B=1$.

Sophisticated calculations of $B$ appeared in the literature using
quark and bag models, lattice calculations and QCD sum rule techniques.

The latter start from a 3-point function involving two pseudoscalar
currents in addition to the $\Delta S$ $(\Delta B)=2$ four quark
operator

\begin{equation}
A(p,p')(p.p')=i^{2}\iint dxdye^{ipx-ip'y}\left\langle 0\left|Tj_{5}(x)\Theta_{_{\Delta S.B=2}}(0)j_{5}(y)\right|0\right\rangle 
\end{equation}

Dispersion relations are written for this quantity and intermediate
states inserted. The sought for matrix elements are provided by the
meson poles. In addition there is a potentially large contribution
arising from the pseudoscalar continuum of which not much is known.

The aim of any reliable calculation is to minimize this contribution
before neglecting it.

In the case of the K-mesons the contribution of the continuum is damped
by use of the Borel (Laplace) transform in which case the damping
is provided by an exponential kernel $e^{\frac{-t}{M^{2}}}$. If the
parameter $M^{2}$, the square of the Borel mass , is small the damping
is good but the contribution of unknown higher order condensates increases
rapidly. If $M^{2}$ increases the contribution of the higher order
condensates decreases but the damping in the resonance region worsens.
An intermediate value of $M^{2}$ has to be chosen. Because $M^{2}$
is an unphysical parameter the results should not depend on it in
a relatively broad interval which is often not the case. The choice
of the parameter which signals the onset of perturbative QCD is another
source of uncertainty.

In the case of the B-mesons, Koerner et al. \cite{Koerner} use inverse
powers of the dispersion variables (moments) which is legitimate because
the matrix elements are infrared safe. As usual the potentially large
contribution of the hadronic continuum is unknown. They estimate it
by using gaps and other ill-known quantities. Recent work on B-meson
mixing using Heavy Quark Effective Theory is also available \cite{KLR}

In this work the aim is to practically eliminate the contribution
of the hadronic continua by introducing kernels which vanish at the
position of the resonances and are very small in a broad region around
them.

The method will also be applied to the evaluation of the quark-gluon
mixed condensate $\left\langle QGC\right\rangle =m_{0}^{2}\left\langle 0\left|\bar{q}q\right|0\right\rangle $
which enters in the calculation and to the evaluation of the K-meson
decay constant.

\section{$\bar{K^{0}}-K^{0}$ Mixing}

In the Standard Model (SM) \cite{Multi2} the mixing of the two eigenstates
of strangeness is predicted as a higher order process which contributes
to the $K_{L}-K_{S}$ mass difference through the so called $\Delta S=2$
box diagram.

The $K_{L}-K_{S}$ mass difference $\Delta m$ is a sum of a long
distance dispersive contribution $\Delta m_{L}$ and a short distance
one $\Delta m_{S}$ proportional to the matrix element $\left\langle \bar{K^{0}}(p')\left|\Theta_{\Delta S=-2}\right|K^{0}(p)\right\rangle $

With
\begin{equation}
\Theta_{_{\Delta S=-2}}=(\bar{s}\gamma_{\mu}(1-\gamma_{5})d)(\bar{s}\gamma_{\mu}(1-\gamma_{5})d)
\end{equation}

Neglecting anomalous dimension factors the parameter $B$ is defined

\begin{equation}
\left\langle \bar{K^{0}}(p')\left|\Theta_{\Delta S=2}\right|K^{0}(p)\right\rangle =\frac{16}{3}Bf_{K}^{2}(p.p')
\end{equation}
$B=1$ in vacuum saturation and $f_{K}=.114\,GeV$

Sophisticated calculations of $B$ followed using quark and bag models,
lattice calculations and QCD sum rules techniques. Unfortunately no
single value for B has emerged.

Start with a 3-point function involving two pseudoscalar currents
in addition to the $\Delta S=2$ four quark operator

\begin{equation}
A(p,p')(p.p')=i^{2}\int\int dxdye^{ipx-ip'y}\left\langle 0\left|Tj_{5}(x)\Theta_{\Delta S=2}(0)j_{5}(y)\right|0\right\rangle 
\end{equation}
where $j_{5}(x)=\bar{d}(x)i\gamma_{5}s(x)$ is the pseudoscalar current

Dispersion relations for this quantity are written and intermediate
states inserted. The $K$-meson poles carry the sought for information
in addition there is the contribution of the strange pseudoscalar
continuum of which not much is known except that it is dominated by
two radial excitations of the $K$, $K(1460)$ and $K(1830)$. In
order to damp the unknown contribution of the continuum Borel (Laplace)
transforms have been used in which case the damping is provided by
an exponential kernel. As discussed in the introduction I shall proceed
otherwise in order to avoid the arbitrariness and instability inherent
to this method. In this work I shall use polynomial kernels in order
to eliminate the contribution of the unknown continuum. The coefficients
of these polynomials are chosen to make the roots coincide with the
masses of the radial excitations of the K.

The amplitude $A(t=p^{2},t'=p'^{2},p.p')$ will be studied at fixed
$p.p'$ and will be denoted by $A(t,t')$

$A(t,t')$ possesses a double pole, two single poles and cuts on the
real $t$, $t'$ axes extending from $th=(m_{K}+2m_{\pi})^{2}$to
infinity stemming from the strange pseudoscalar intermediate states.

\begin{equation}
A(t,t')(p.p')=\frac{2f_{K}^{2}m_{K}^{4}\left\langle K^{0}\left|\Theta_{\Delta S=2}\right|K^{0}\right\rangle }{(m_{s}+m_{d})^{2}(t-m_{K}^{2})(t'-m_{K}^{2})}+\frac{\varPhi(t)}{t'-m_{K}^{2}}+\frac{\varPhi(t')}{t-m_{K}^{2}}+\cdots
\end{equation}

Consider now the double integral in the complex $t$ and $t'$ planes

\begin{equation}
\frac{1}{(2\pi i)^{2}}\int_{c}\int_{c'}dtdt'P(t)P(t')A(t,t')(p.p')
\end{equation}
where $c$ and $c'$ are the contours shown on Fig. 1, $f_{K}$ is
the $K$ decay constant and $P(t)$ is a so far arbitrary entire function.

\begin{figure}[H]
\centering{}\includegraphics[width=11cm]{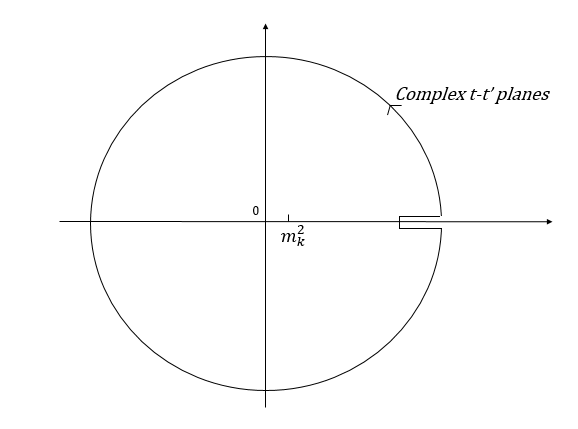}\caption{The contours of integration c,c'}
\end{figure}

Because $\Phi(t),\Phi(t')$ have no singularities inside the contours
of integration the single poles do not contribute to the double integral
and we are left with

\begin{equation}
\frac{2f_{K}^{2}m_{K}^{4}}{(m_{s}+m_{_{d}})^{2}}\left\langle \bar{K^{0}}\left|\Theta_{\Delta S=2}\right|K^{0}\right\rangle P^{2}(m_{K}^{^{2}})=\frac{1}{(2\pi i)^{2}}\int_{c}\int_{c'}dtdt'P(t)P(t')A(t,t')(p.p')
\end{equation}

The integrals over the cuts represent the contribution of the pseudoscalar
strange continuum. $P(t)$ is now chosen to be a second order polynomial
whose roots coincide with the masses squared of the radial excitations
of the $K$, $K(1400)$ and $K(1870)$.

\begin{equation}
P(t)=1-a_{1}t-a_{2}t^{2}=1-.768GeV^{-2}t+.14GeV^{-4}t^{2}
\end{equation}

This choice of $P(t)$ and $P(t')$ which vanishes at the radial excitations
of the K and is very small in a very broad region around them, practically
eliminates the contribution of the hadronic continuum and leaves us
with the integrals on the circles of large radius $R$ where $A(t,t')$
can be replaced by $A^{QCD}(t,t')$ so that using $\left\langle \bar{K^{0}}\left|\Theta_{\Delta S=2}\right|K^{0}\right\rangle =\frac{16}{3}f_{K}^{2}B(p.p')$
gives

\begin{equation}
\frac{32}{3}\frac{f_{K}^{4}m_{K}^{4}}{(m_{s}+m_{d})^{2}}P^{2}(m_{K}^{2})B=\frac{1}{(2\pi i)^{2}}\ointop\ointop dtdt'P(t)P(t')A^{QCD}(t,t')\label{eq:11}
\end{equation}

$A^{QCD}(t,t')$ is the sum of a factorizable and a non-factorizable
part \cite{Multi1}

\begin{equation}
A^{QCD}=A_{f}^{QCD}+A_{nf}^{QCD}
\end{equation}
where
\begin{equation}
A_{f}^{QCD}=\frac{8}{3}\Pi_{5}(t)\varPi_{5}(t')\label{eq:13}
\end{equation}

\begin{equation}
\varPi_{5}(t)=-\frac{3}{8\pi^{2}}m_{s}ln(-t)+\frac{\left\langle \bar{d}d+\bar{s}s\right\rangle }{t}+\frac{m_{s}\left\langle a_{s}GG\right\rangle }{8t^{2}}+\frac{0}{t^{3}}+\cdots\label{eq:14}
\end{equation}
and

\begin{equation}
A_{nf}^{QCD}=\frac{2}{3}m_{0}^{2}\left\langle \bar{q}q\right\rangle ^{2}(\frac{1}{t^{2}t'}+\frac{1}{t'^{2}t})+\frac{1}{4\pi^{2}}m_{0}^{2}\left\langle m_{s}\bar{q}q\right\rangle \frac{1}{tt'}(\ln(\frac{-t}{\mu^{2}})+\ln(\frac{-t}{\mu^{2}}))-[\frac{4\pi^{2}}{9}\left\langle \bar{q}q\right\rangle ^{2}\left\langle a_{s}GG\right\rangle +\frac{13}{288}m_{0}^{4}\left\langle \bar{q}q\right\rangle ^{2}]\frac{1}{t^{2}t'^{2}}+...
\end{equation}

Because $B_{f}=1$, eqs. (\ref{eq:11}), (\ref{eq:13}) and (\ref{eq:14})
yield

\begin{equation}
\frac{2f_{K}^{2}m_{K}^{2}}{m_{s}+m_{d}}P(m_{K}^{2})=I\label{eq:16}
\end{equation}
where

\begin{align}
I & =\frac{1}{2\pi i}\ointop dtP(t)\Pi_{5}(t)\\
 & =-\frac{3m_{s}}{8\pi^{2}}\frac{1}{2\pi i}\ointop dtP(t)\ln(-t)+\left\langle \bar{d}d+\bar{s}s\right\rangle +\frac{a_{1}m_{s}}{8}\left\langle a_{s}GG\right\rangle \nonumber 
\end{align}

Because $ln(-t)$ has a cut on the positive t-axis which starts at
the origin the integral over the circle in the equation above can
be transformed into an integral over the real axis so that

\begin{equation}
I=-\frac{3m_{s}}{8\pi^{2}}\int_{0}^{R}dtP(t)+\left\langle \bar{d}d+\bar{s}s\right\rangle +\frac{a_{1}m_{_{s}}}{8}\left\langle a_{s}GG\right\rangle \label{eq:18}
\end{equation}

The choice of R is determined by stability considerations. It should
not be too small as this would invalidate the Operator Product Expansion
on the circle, nor should it be too large because $P(t)$ would start
enhancing the contribution of the continuum instead of suppressing
it. We seek an intermediate range of $R$ for which the integral in
eq. (\ref{eq:18}) is stable.

The integral $i(R)=\int_{0}^{R}dtP(t)$ is seen to be stable for $2GeV^{2}\preceq R\preceq4GeV^{2}$,
$i(R)\eqsim.83GeV$ as shown in Fig. 2.

\begin{figure}[H]
\begin{centering}
\includegraphics[width=14cm]{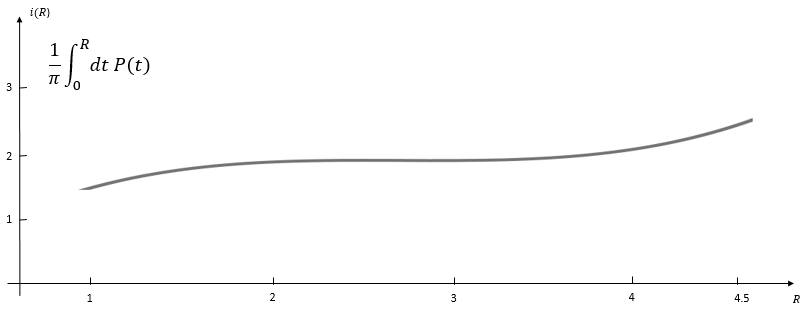}
\par\end{centering}
\caption{The variation of $i(R)=\int_{0}^{R}dtP(t)$ as a function of $R$
in $Gev$}
\end{figure}

Then

\begin{equation}
I=\left\langle \bar{d}d+\bar{s}s\right\rangle +\frac{3}{8\pi^{2}}m_{s}i(R)+\frac{a_{1}m_{s}}{8}\left\langle a_{s}GG\right\rangle 
\end{equation}

so that

\begin{equation}
\frac{2f_{K}^{2}m_{K}^{2}}{m_{s}+m_{d}}P(m_{K}^{2})=\left\langle \bar{d}d+\bar{s}s\right\rangle +\frac{3}{8\pi^{2}}m_{s}i(R)+\frac{a_{1}m_{s}}{8}\left\langle a_{s}GG\right\rangle \label{eq:20}
\end{equation}

The equation above is dominated by the quark condensate term $\left\langle \bar{d}d+\bar{s}s\right\rangle $.
Eq. (\ref{eq:20}) is seen to be a version of the Gell-Mann, Oakes,
Renner relation \cite{GOR} in the strange sector modified by $SU(3)\times SU(3)$
chiral symmetry breaking.

Turn now to the contribution of the non-factorizable part. Similar
manipulations lead to

\begin{equation}
\frac{4f_{K}^{4}m_{K}^{4}P^{2}(m_{K}^{2})B_{nf}}{(m_{s}+m_{d})^{2}}=-\frac{1}{2}a_{1}m_{0}^{2}\left\langle \bar{q}q\right\rangle ^{2}+\frac{3m_{0}^{2}}{16\pi^{2}}\left\langle m_{s}\bar{q}q\right\rangle \ln\frac{R}{\mu^{2}}-\frac{3}{8}(\frac{4\pi^{2}}{9}\left\langle \bar{q}q\right\rangle ^{2}\left\langle a_{s}GG\right\rangle +\frac{13}{288}m_{0}^{4}\left\langle \bar{q}q\right\rangle ^{2})a_{1}^{2}
\end{equation}

Values of $m_{0}^{2}$, which parameterizes the quark-gluon mixed
condensate vary over a large range in the literature. The method presented
here offers an independent evaluation of this quantity :

The integral

\begin{equation}
\frac{1}{(2\pi i)^{2}}\int_{c}dtP(t)(t-m_{K}^{2})\int_{c'}dt'P(t')(t'-m_{K}^{2})A^{QCD}(t,t')=I'_{f}+I'_{nf}=0
\end{equation}

vanishes because the singularities inside $c$ and $c'$ have been
removed .

\begin{equation}
I'_{f}=\frac{8}{3}I'^{2}
\end{equation}

\begin{align}
I' & =\frac{1}{(2\pi i)}\ointop dtP(t)(t-m_{K}^{2})\{-\frac{3m_{s}^{2}}{8\pi^{^{2}}}\ln(-t)+\frac{\left\langle \bar{d}d+\bar{s}s\right\rangle }{t}-\frac{m_{_{s}}\left\langle a_{s}GG\right\rangle }{8t^{^{2}}}\}\\
 & =-m_{K}^{2}\left\langle \bar{d}d+\bar{s}s\right\rangle -\frac{3m_{s}}{8\pi^{2}}i'(R)-\frac{m_{s}}{8}\left\langle a_{s}GG\right\rangle (1+a_{1}m_{K}^{2})
\end{align}

where $i'(R)=\int_{0}^{R}dtP(t)(t-m_{K}^{2})$

The non-factoizable contribution is

\begin{equation}
I'_{nf}=-\frac{4}{3}m_{0}^{2}\left\langle \bar{q}q\right\rangle ^{2}m_{K}^{2}(1+a_{1}m_{K}^{2})-\frac{m_{s}}{2\pi^{2}}m_{0}^{2}\left\langle \bar{q}q\right\rangle m_{K}^{2}[i''(R)-m_{K}^{2}\ln\frac{R}{\mu^{2}}]-[\frac{4\pi^{2}}{9}\left\langle \bar{q}q\right\rangle ^{^{2}}\left\langle a_{sGG}\right\rangle +\frac{13}{288}m_{0}^{4}\left\langle \bar{q}q\right\rangle ^{2}](1+a_{1}m_{K}^{2})^{2}
\end{equation}

where $i''(R)=\int_{0}^{R}dt[1+a_{1}m_{K}^{2}-(a_{1}-a_{2}m_{K}^{2})t-a_{2}t^{2}]$

The condensate $\left\langle \bar{d}d+\bar{s}s\right\rangle $ dominates
our equations. It could be obtained from eq. (\ref{eq:16}), an improved
calculation (to five loops) is found in \cite{DNS} it gives

\begin{equation}
-(m_{s}+m_{d})\left\langle \bar{d}d+\bar{s}s\right\rangle =(.39\pm.03).10^{-2}\,GeV^{4}
\end{equation}

This,with $(m_{s}+m_{d})=(108\pm8)\,MeV$yields $m_{0}^{2}\eqsim1.0\,GeV^{2}$
which determines $B_{nf}$

\begin{equation}
B_{nf}=-.09\ or\ B\eqsim.91
\end{equation}

\section{$f_{K}$ to Five Loops}

Theoretical calculations of the weak decay constants $f_{K}$ and
$f_{\pi}$ are of great interest. This has been done recently in the
context of the extended Nambu-Jona-Lasinio model \cite{Volkov}, using
an improved holographic wave function \cite{Chang} in the light-front
quark model \cite{Choi} or on lattice calculations \cite{Carras}.

I offer here instead a QCD calculation of $f_{K}$ to five loops.
Start with the correlator

\begin{equation}
\Pi_{\mu\nu}(t=q^{2})=i\intop dxe^{iqx}\left\langle 0\left|TA_{\mu}(x)A_{\nu}(0)\right|0\right\rangle =(q_{\mu}q_{\nu}-g_{\mu}q^{2})\Pi^{1}(t)+q_{\mu}q_{\nu}\Pi^{0}(t)
\end{equation}

Let $\Pi(t)=\Pi^{0+1}(t)$ and consider

\begin{equation}
\int_{c}dtP(t)\Pi(t)=f_{K}^{2}P(m_{K}^{2})=\frac{1}{\pi}\intop_{0}^{R}dtP(t)Im\Pi(t)+\varoint dtP(t)\Pi^{^{QCD}}(t)
\end{equation}

As before the polynomial $P(t)$ is chosen in order to eliminate the
contribution of the integral on the cut. We have now to take into
account the axial-vector resonances in addition to the pseudoscalar
ones, i.e. $K_{1}(1273)$ and $K_{1}(1402)$ in addition to $K(1460)$
and $K(1830)$.

The choice $P(t)=1-1.42t+.648t^{2}-.093t^{3}$with the coefficients
in appropriate powers of $GeV$ achieves the purpose of eliminating
the contribution of the continuum. Here

\begin{equation}
\Pi^{QCD}(t)=\Pi_{pert}(t)+\frac{c_{1}}{t}+\frac{c_{2}}{t^{2}}+\frac{c_{3}}{t^{3}}+\cdots
\end{equation}

\begin{equation}
4\pi Im\Pi_{pert}(t)=1+a_{s}(r)+a_{s}^{2}(r)l_{2}(t,r)+a_{s}^{3}(r)l_{3}(t,r)+a_{s}^{4}(r)l_{4}(t,r)
\end{equation}

The $l_{i}(t,r)$ and the strong coupling constant $a_{s}(r)$ are
known to 5-loop order \cite{CKS,BCK} and the non perturbative condensates
are given in \cite{BNP}

\begin{align}
c_{1} & =\frac{3m_{s}^{2}}{4\pi^{2}}(1+\frac{7}{3}a_{s})\nonumber \\
c_{2} & =\frac{1}{12}(1-\frac{11}{18}a_{s})\left\langle a_{s}GG\right\rangle +(1-\frac{a_{s}}{3})\left\langle m_{s}\bar{s}s\right\rangle +O(a_{s}^{2})\\
c_{3} & =-a_{s}\frac{32\pi^{2}}{9}[\left\langle \bar{q}q\right\rangle \left\langle \bar{s}s\right\rangle +\frac{1}{9}\left\langle \bar{q}q\right\rangle ^{2}]+O(a_{s}^{2})\nonumber 
\end{align}

The integral of $\Pi_{pert}(t)$ over the circle is transformed into
an integral over the cut once again and finally

\begin{equation}
2f_{K}^{2}P(m_{K}^{^{2}})=\frac{1}{\pi}\int_{0}^{R}dtP(t)Im\Pi_{pert}(t)+c_{1}-a_{1}c_{2}-a_{2}c_{3}
\end{equation}

With the standard values $m_{s}=.10GeV^{4}$, $\left\langle a_{s}GG\right\rangle =.013GeV^{4}$,
$\left\langle \bar{s}s\right\rangle =.6\left\langle \bar{q}q\right\rangle $,
$\left\langle \bar{q}q\right\rangle =.02GeV^{3},$the final result
is

\begin{equation}
f_{K}=.107\,GeV
\end{equation}

The pion decay constant $f_{\pi}$ could also be studied. In this
case data on the continuum is available from $\tau$ decay and yields
$f_{\pi}$ \cite{DNS2}. The method used above can likewise be applied
taking into account the pseudoscalar and axial-vector resonances with
the result 
\begin{equation}
f_{\pi}=.092\,GeV
\end{equation}

\section{$\bar{B^{0}}-B^{0}$ Mixing}

Turn now to $\bar{B^{0}}-B^{0}$ mixing. Start with the three point
correlation function

\begin{equation}
\Pi(p_{1},p_{2})=\int\int dxdye^{i(p_{1}x-p_{2}y)}\left\langle 0\left|Tj_{B}(x)O(0)j_{b}(y)\right|0\right\rangle 
\end{equation}

The operator $j_{B}=(m_{b}+m_{d})di\gamma_{5}b=\partial_{\mu}(d_{L}\gamma_{\mu}\gamma_{5}b_{L})$
is the interpolating current for the $B$ meson and $\left\langle 0\left|j_{B}(0)\right|B^{0}\right\rangle =f_{B}m_{B}^{2}$

The relevant quantity to calculate is the matrix element $A=\left\langle \bar{B^{0}}\left|O(\mu)\right|B^{0}\right\rangle $
where $O(\mu)=(b_{L}\gamma_{\mu}d_{L})(d_{L}\gamma_{\mu}b_{L})$ is
the local 4-quark operator at the normalization point $\mu$ which
can be used to evaluate the splitting of heavy and light mass eigenstates.
The simplest approach (factorization) \cite{Multi2} reduces $A$
to

\begin{equation}
A^{f}=\frac{8}{3}\left\langle \bar{B^{0}}\left|\bar{b}_{L}\gamma_{\mu}d_{L}\right|0\right\rangle \left\langle 0\left|\bar{b}_{L}\gamma_{\mu}d_{L}\right|B^{0}\right\rangle =\frac{2}{3}f_{B}^{2}m_{B}^{2}
\end{equation}

where $\left\langle 0\left|\bar{b}_{L}\gamma_{\mu}\gamma_{5}d_{L}\right|B^{0}(p)\right\rangle =if_{B}p_{\mu}$

The deviation from factorization is again parametrized by $B_{B}$
defined as $A=B_{B}A^{f}$. In factorization $B_{B}=1$.

A nice way to calculate $B$, using inverse moments, was used in \cite{Koerner}.
In the present work I shall follow their approach but the contribution
of the higher resonances and continuum shall be estimated in a different
less model dependent and more reliable way.

A dispersion representation of the correlator reads

\begin{equation}
\Pi(p_{1}^{2},p_{2}^{2},q^{2})=\iint dt_{1}dt_{2}\frac{\varrho(t_{1},t_{2},q^{2})}{(t_{1}-p_{1}^{2})(t_{2}-p_{2}^{2})}
\end{equation}

where $q=p_{1}-p_{2}$. Consider the moments of the correlation function
at $p_{1}^{2}=p_{2}^{2}=q^{2}=0$

\begin{equation}
M_{ij}=\iint dt_{1}dt_{2}\frac{\varrho(t_{1},t_{2},0)}{t_{1}^{i}t_{2}^{j}}\label{eq:40}
\end{equation}

Because the origin is infrared safe $M_{ij}$ can be computed in QCD
\cite{SVZ}. It also has a phenomenoligical representation

\begin{equation}
M_{ij}^{^{ph}}=\frac{8}{3}B_{B}f_{B}^{4}\frac{m_{B}^{2}}{m_{B}^{2(i+j)}}+\cdots
\end{equation}

where the ellipses stand for the contribution of the higher resonances
and the continuum. Separating the factorizable part we obtain two
sum rules

\begin{align}
\frac{8}{3}B_{B}f_{B}^{4}m_{B}^{2}m_{B}^{2(i+j)}+\Delta_{ij} & =M_{ij}\nonumber \\
\frac{8}{3}f_{B}^{4}m_{B}^{2}m_{B}^{2(i+j)}+\Delta_{ij}^{'f} & =M_{ij}^{f}\label{eq:42}
\end{align}

The l.h.s of the above eqs. represent the phenomenology and the r.h.s
the QCD theoretical expressions.

The $\Delta'$s are the contributions of the higher resonances and
the continuum. While \cite{Koerner} try to estimate their contribution
using gaps and other ill-known parameters, I aim to eliminate it altogether.

For this purpose note that the integrals in eq. (\ref{eq:40}) are
fast convergent so that the bulk of the contribution to the continuum
come from the vicinity of the first resonance. If this contribution
is eliminated $\Delta_{ij}$ and $\Delta'_{ij}$ become negligible
and can be discarded. This is done by using instead of eq. (\ref{eq:40})

\begin{align}
I_{ij} & =\iint dt_{1}dt_{2}\frac{\varrho(t_{1},t_{2})}{t_{1}^{i}t_{2}^{j}}(\frac{m'^{2}}{t_{1}}-1)(\frac{m'^{2}}{t_{2}}-1)\nonumber \\
 & =M_{ij}-m'^{2}M_{i,j+1}-m'^{2}M_{i+1,j}+m'^{4}M_{i+1,j+1}
\end{align}

If $m'$ is close to the mass of the first resonance, the factor ($\frac{m'^{2}}{t}$
-1) annihilates the integrand in its vicinity. Because of the fast
convergence due to the denominators the main contribution of the continuum
is eliminated and this justifies the neglect of $\Delta$ and $\Delta'$.

Taking $i=j=2$ eqs. (\ref{eq:42}) become

\begin{align}
\frac{8}{3}\frac{B_{B}f_{B}^{4}m_{B}^{2}}{m_{B}^{2(i+j)}}(\frac{m'^{2}}{m_{B}^{2}}-1)^{2} & =I_{ij}\nonumber \\
\frac{8}{3}\frac{f_{B}^{4}m_{B}^{2}}{m_{B}^{2(i+j)}}(\frac{m'^{2}}{m_{B}^{2}}-1)^{2} & =I_{ij}^{f}
\end{align}

or

\begin{equation}
B_{B}=\frac{M_{22}-2m'^{2}M_{23}+m'^{4}M_{33}}{M_{22}^{f}-2m'^{2}M_{23}^{f}+m'^{4}M_{33}^{f}}
\end{equation}

The $M$s are the theoretical QCD values and the $M^{f}$ their factorizable
counterparts.

It was found in \cite{Koerner} that

\begin{equation}
M_{ij}=\frac{m_{b}^{6}a_{ij}}{m_{b}^{2(i+j)}}(1+\frac{a_{s}}{4}(b_{ij}^{f}+b_{ij}^{nf}))+M_{ij}^{non-perturbative}
\end{equation}

Expressions for $M_{ij}^{non-perturbative}$ are given in \cite{Pivov}.
In the present approach their contribution vanishes. The quantities
$a_{ij}$, $b_{ij}^{f}$ , $b_{ij}^{nf}$ represent the LO, NLO and
non-factorizable contributions.

Then, if

\begin{equation}
B_{B}=\frac{I_{22}}{I_{22}^{f}}=1+\frac{a_{_{s}}}{4}\delta
\end{equation}

\begin{equation}
\delta=\frac{a_{22}b_{22}^{nf}-2(\frac{m'}{m_{b}})^{2}a_{23}b_{23}^{nf}+(\frac{m'}{m_{b}})^{4}a_{33}b_{33}^{nf}}{a_{22}-2(\frac{m'}{m_{b}})^{2}a_{23}+(\frac{m'}{m_{b}})^{4}a_{33}}
\end{equation}

The corresponding expression in the work of \cite{Koerner} is

\begin{equation}
\delta=b_{22}^{nf}+\delta R+\delta C
\end{equation}

Where $\delta R$ and $\delta C$ are parameters which account for
the resonances and continuum contribution and which they estimate
by fitting and using gap parameters. The present approach avoids this
arbitrariness.

The $a_{ij}$ are 

\begin{equation}
a_{ij}=m_{b}^{2(i+j)-6}M_{ij}^{LO}
\end{equation}

\begin{equation}
M_{ij}^{LO}=\iint\frac{dt_{1}dt_{2}}{t_{1}^{i},t_{2}^{j}}\frac{4}{3}(t_{1}+t_{2})\varrho(t_{1})\varrho(t_{2})
\end{equation}

\begin{equation}
\varrho(t)=\frac{3}{16\pi^{2}}m_{q}^{2}(1-\frac{m_{q}^{2}}{t})
\end{equation}

which yields

\begin{equation}
a_{22}=\frac{1}{(16\pi^{2})^{2}}\frac{8}{3},\quad a_{23}=\frac{1}{(16\pi^{2})^{2}}\frac{2}{3},\quad a_{33}=\frac{1}{(16\pi^{2})^{2}}\frac{1}{6}
\end{equation}

The $b_{ij}^{nf}$ are given in \cite{Koerner}

\begin{equation}
b_{22}^{nf}=.68,\quad b_{23}^{nf}=1.22,\quad b_{33}^{nf}=1.96
\end{equation}

The Particle Data Group lists two candidates for $m'$, $m'(5.84)$
and $m'(5.97)$. A reasonable choice is then $(\frac{m'}{m_{b}})^{^{2}}=2.0$
which yields $\frac{a_{s}}{4}\delta=-.006$ or $B_{B}\simeq1.0$

Deviation from factorization is negligible.

\section{Discussion}

In this work i have studied the contribution of the hadronic continuum
to deviations from factorization in neutral K and B meson mixing.
In both cases I minimized this contribution before neglecting it by
using simple kernels in the dispersion integrals which vanish at the
low lying resonances, in the case of the K-meson a polynomial kernel
was used, the same method (and kernel ) was also used in the calculation
of the quark- gluon mixed condensate as well as the evaluation of
the K-meson decay coupling constant $f_{K}$.

For the B-meson inverse moments were used. These render the integrals
fast convergent and concentrate the contribution to the dispersion
integral in the vicinity of the first resonance. In this case the
kernel used is of the form $(\frac{m'^{2}}{t}-1)$ where $m'$ lies
in the vicinity of the first resonance.

This method avoids the arbitrariness and instability inherent to previously
used ones.

\pagebreak{}

\end{document}